\documentclass[twocolumn,preprintnumbers,amsmath,amssymb]{revtex4}

\usepackage{graphicx}
\usepackage{dcolumn}
\usepackage{bm}

\begin{document}

\title{Localized heating in nanoscale Pt constrictions measured using blackbody radiation emission}

\author{Daniel R. Ward$^{1}$}
\author{Naomi J. Halas$^{2,3}$}
\author{Douglas Natelson$^{1,3}$}
\affiliation{Department of Physics and Astronomy$^1$,  Department of Chemistry$^2$,  Department of Electrical and Computer Engineering$^3$, Rice University, 6100 Main Street, Houston, Texas 77005}

\date{\today}

\begin{abstract}
Using thermal emission microscopy, we investigate heating in Pt nanowires before and during electromigration.  The wires are observed to reach temperatures in excess of 1000~K.  This is beyond the thermal decomposition threshold for many organic molecules of interest for single molecule measurements with electromigrated nanogaps.  Blackbody spectra of the hot Pt wires are measured and found to agree well with finite element modeling simulations of the electrical and thermal transport.  
\end{abstract}


\maketitle

Electromigration has emerged as a common technique for the fabrication
of nanometer-scale gaps between conducting
electrodes\cite{ParketAl99APL,strachanetAl2008PRL,TaychatanapatetAl07NL,heerscheelAl07APL}.
Electromigration is driven by momentum transfer between
conduction electrons and metal atoms at surfaces and grain boundaries.
Taking advantage of this process, a large current density may be
applied to constriction in a lithographically patterned metal wire,
breaking the wire into nano-spaced source and drain electrodes.  Local
Joule heating assists electromigration by enhancing
atomic diffusion.

The resulting electrodes are often used for studies of single-molecule
electronic conduction\cite{ParketAl00Nature,NatelsonetAl06CP}, with
electromigration performed in the presence of molecules of interest.
Because of concerns about molecular stability at elevated
temperatures, there is much interest in the magnitude of Joule heating
during the electromigration process.  In particular, high temperatures
($>$600~K) can induce thermal break down in complex molecules and
provide sufficient energy to remove weakly bound molecules from the
electrode surfaces.  Recently the temperature in these and similar
nanostructures has been assessed via the observation of melting of
nanoparticles decorating the metal
surface\cite{TaychatanapatetAl07NL,BrintlingeretAl08NL}.  However,
this requires special surface preparation and high vacuum electron
microscopy, and has a temperature resolution limited by the melting
point range of the nanoparticles.

In this letter we instead employ a variation of thermal emission
microscopy\cite{IppolitoetAl04APL,DhokkaretAl07MST} to infer
temperature profiles in nanostructures on a length scale of
approximately 1~$\mu$m.  Specifically we examine Pt constrictions
using blackbody radiation as a simple method to determining the local
heating.  The Pt nanowires are inferred to reach very high
temperatures exceeding 1000~K before the onset of electromigration.
Optical measurements show good agreement with finite element modeling
(FEM) of the heating process.  With an appropriate infrared detection
system this approach may be applied to other metals without special
surface treatments.  The results for Pt nanostructures indicate a
significant challenge in creating Pt electromigrated gaps suitable for
use in single molecule electronic devices.

\begin{figure}[h!]
\includegraphics[width=8cm]{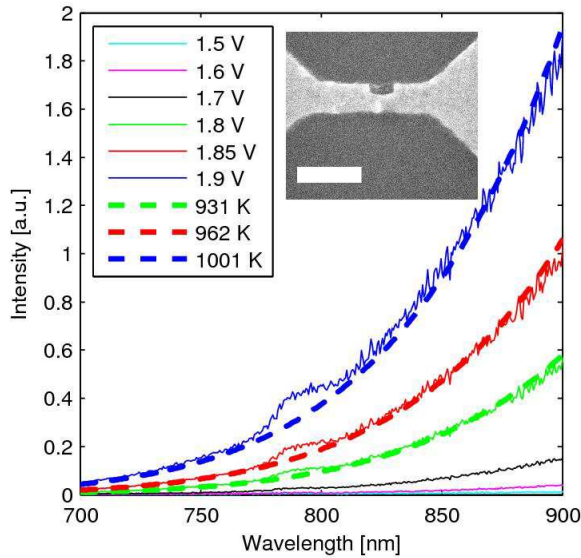}
\caption{\label{figLightEmission}
Observed light emission after calibration correction from a 1030~nm long 180~nm wide Pt bowtie at various biases.  The three dashed lines indicate theoretical blackbody curves at 931~K, 963~K, and 1001~K which show good agreement with the measurements at 1.80~V (3.32~mA), 1.85~V (3.38~mA) and 1.90~V (3.44~mA) applied bias (current), respectively (color online).  
Inset: SEM image of a typical Pt wire constriction, 530~nm long and 140~nm wide.  The scale bar is 300~nm in length.   The notch in the middle of the wire is the result of partially completed electromigration.  In this device current flow was from left to right.
}
\end{figure}

Platinum nanowires with widths of approximately 140~nm and lengths ranging from 200~nm to 1200~nm were fabricated using e-beam lithography on silicon substrates with 200~nm of thermal oxide.  Each nanowire was connected via 6~$\mu$m wide wires to large contact pads for electrical probing.  A 1~nm Ti adhesion layer and 15~nm Pt film were deposited using an e-beam evaporator at rates of 1~\AA/s and 0.5~\AA/s respectively.  Lift off was performed in acetone after which samples were O$_2$ plasma cleaned for one minute to remove any organic residue.  A typical device is picture in the inset to Figure~\ref{figLightEmission}.

The blackbody spectrum of heated Pt nanowires was measured in air at room temperature in a confocal Raman microscope with a silicon CCD detector.  Samples were heated in-situ by applying a bias across the sample using electrical probes.  The microscope was aligned over the nanowire structure by positioning the Raman laser spot over the center of the nanowire.  Blackbody spectra were collected at fixed bias values with the laser off.  Each spectrum was measured by averaging two 15~s integrations.  Spatial mappings of the light emission were also acquired by rastering the stage with the nanowire under the microscope.  Spatial map spectra were integrated for one to two seconds at each position.  Additional spatial maps were taken while using the 785~nm laser to measure the Raman response of the Si substrate.  The strong 520~cm$^{-1}$ Si Raman peak is attenuated when the Pt film is over the Si substrate, allowing the physical layout of the nanowire to be determined from the map of the Si peak intensity.   The combined Raman/blackbody measurements allow us to map the location of the light emission to the physical position along the wire.  

The absolute temperature of the nanowires was determined by fitting the measured spectra to theoretical blackbody emittance per wavelength of the form 
$M=\frac{2 \pi c}{\lambda^4 (e^{\frac{hc}{\lambda k T}}-1)}$
where M has units of photons/(cm$^2$-s-$\mu$m). The efficiency of the Si CCD in our microscope drops off rapidly at wavelengths exceeding 1~$\mu$m.  To account for this loss of efficiency the output of the CCD was calibrated using a known 2960~K blackbody source.  The blackbody source was imaged by diffusing the source's light with spectralon glass and imaging the surface of the glass.  This ensures that the calibration takes into consideration all the losses of the optics along with the CCD efficiency.

Figure~\ref{figLightEmission} shows the measured blackbody spectrum after correction for a Pt wire device 1030~nm long and 180~nm wide at different bias voltages.  Approximately half the applied voltage is dropped across the wire with the rest dropped in the connections from the contact pads to the wire.  The current at 1.90~V was 3.44 mA yielding a current density of $1.3 \times 10^{12}$~A/m$^2$. The temperature at the center of each wire is determined by fitting the calibration-corrected spectrum measured at the center of the wire to a perfect blackbody spectrum.  The fit has two free parameters, temperature and amplitude; the latter is a measure of how efficiently the microscope can gather and detect photons.  Fits are performed over the wavelength range of 650~nm to 950~nm with the region from 770~nm to 805~nm removed to avoid the spectral region affected by the 785~nm holographic beam splitter used for Raman measurements.  Due to the exponential decrease in photon counts as temperature is reduced, fits for temperatures below 900~K yield increasingly poor results due to inadequate signal-to-noise.  For instance in Figure~\ref{figLightEmission}B, for temperatures less than 900~K the signal-to-noise ratio is less than one.  To assist in analysis we fit the data obtained at the highest voltage first, and then fix the amplitude parameter in fits for the same wire at lower biases.  This is a valid method as the amplitude is a measure of the optical system's transmission efficiency and should remain constant as long as the microscope's configuration and sample position in the focal region remain unchanged.

We performed FEM simulations using commercially available software to verify our results.  We model the Pt film using the same dimensions as the actual wires but omitting the large contact pads as they do not provide significant heating due to their relatively low resistance.  The contact pads also do not contribute significantly to heat dissipation as verified in our model.  The substrate was modeled as 200~nm of thermal oxide on top of a 2~$\mu$m thick Si slab, the bottom of which is assumed to be held fixed at room temperature.  The substrate thickness was chosen for computational efficiency; we verified that increasing the thickness from 2~um to 5~um or more has no effect on the heating of the wire.  Finally where appropriate we model heat transfer from the Pt wire into the surrounding air by modeling the heat transfer coefficient of the film to air to be 5~W/(m$^2$~K).  Bulk values are used for all thermal conductivities and for the electrical properties of the oxide and substrate.  We experimentally measured the resistivity of our 15~nm Pt films using the Van der Pauw technique.  Additionally we measured the linear temperature dependence of the resistivity from 200~K to 400~K.  We experimentally found the resistivity to be $3.9 \times 10^{-7} \Omega$~m and the temperature coefficient $\alpha$ to be $7.19 \times 10^{-4}$~1/K where $\alpha$ is defined in the resistivity ($\rho$) equation
$\rho(T)=\rho_{T=T_0}(1+\alpha(T-T_0))$.
For each simulation we record the temperature at the center of the wire at the same sample biases used in our experiment.

\begin{figure}[h!]
\includegraphics[width=8cm]{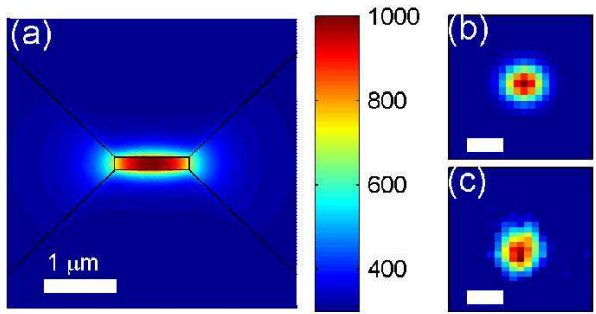}
\caption{\label{figFEMModel}
(color online)(a) Finite element model of temperature profile in a Pt bowtie 1030~nm long and 180~nm wide with a bias of 1.90~V.  Heating has clearly been localized to the nanowire region.
(b) Calculated light emission intensity from modeled bowtie based upon integrated emission from 650~nm to 950~nm assuming a gaussian sensing profile of FWHM 800~nm.  Scale bar indicates 1~$\mu$m.  Color scale ranges from blue (zero intensity) to red (highest intensity).
(c) Experimentally measured light emission intensity from bowtie.  The shape and extent of emission matches predicted emission well.  Scale bar indicates 1~$\mu$m.  Color scale ranges from blue (zero intensity) to red (highest intensity).
}
\end{figure}

\begin{figure}[h!]
\includegraphics[width=8cm]{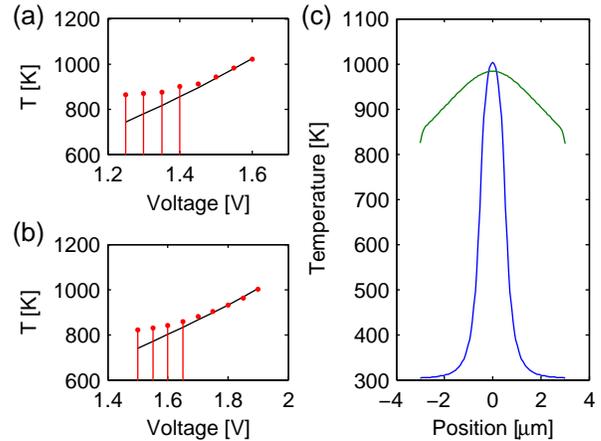}
\caption{\label{figHeating}
(a \& b) Inferred temperature (red points) of 680~nm long 140~nm wide and 1030~nm long 180~nm wide Pt wires as a function of applied bias.  Error bars indicate 95\% confidence interval determined via $\chi ^{2}$ analysis.  For lower voltages the temperature can not be experimentally determined.  Black lines are simulated temperatures based upon FEM.  Inferred and simulated temperatures differ at low temperatures due to the small signal to noise ratio in experimental measurements. 
(c) (color online) FEM-based comparison of the true temperature (blue line) at a given point in 1030~nm long Pt wire biased with 1.90~V versus what the observed temperature (green line) would be based upon the convolution of the 800~nm FWHM microscope sensitivity.  At each spatial coordinate the resulting spectrum is least-squares curve fit in amplitude and temperature to detemine the observed temperature.  
}
\end{figure}

Experimental measurements correlate well with the FEM model.  Figure~\ref{figFEMModel}a is the modeled temperature profile for an actual device.  Figure~\ref{figFEMModel}b is a spatial map of the expected integrated emission intensity from 650~nm to 1.00~$\mu$m.  The emission map is in good agreement with the measured emission for this device as seen in Figure~\ref{figFEMModel}c.  The highest temperatures reported in the model are in agreement with our measured values as shown for two devices in Figure~\ref{figHeating}a,~b.

Small deviations between the experimental and model temperatures may also be the result of the convolution of the microscope's sampling area and the true temperature profile.  We have measured the spatial sensitivity of our microscope, and it can be well approximated by a Gaussian profile with a FWHM of approximately 500~nm.  Figure~\ref{figHeating}c is a 1D calculation to find the effective temperature one would expect to measure at each point along a 1030~nm long Pt wire based upon our sensitivity profile.  The effective temperature is based upon fitting a blackbody spectrum to the sum of the blackbody spectra from each wire element multiplied by the microscope sensitivity at that element.  It should be noted that the actual spectral amplitude drops off significantly when measured off center, as observed in Figure~\ref{figFEMModel}b, but the relative wavelength dependence, which allows determination of the temperature, is unaffected.  This is beneficial as it makes the measured temperature insensitive to small instrumental shifts away from the wire center.  However, these point spread effects also mean the true highest temperature is never measured directly.

The temperatures reached during electromigration in Pt wires are considerably higher than those observed for Au wires\cite{TaychatanapatetAl07NL}.  This is sensible, since the refractory nature of Pt would lead one to expect that higher voltages would be required for migration.   The comparatively extreme temperatures measured are likely too high for making molecular electronic devices without damaging the molecule.  We modeled three possible electromigration scenarios: in air at room temperature, in vacuum at room temperature, and in vacuum with the substrate cooled to 80~K.  For the 1030~nm long wire at 1.90~V little change is observed regardless of conditions.  Air cooling provides only a small temperature drop from 1004~K to 1001~K.  Cooling the substrate has more effect but only reduces the final temperature to 905~K, still above the thermal decomposition threshhold of many organic compounds.

We note that recent measurements of blackbody emission from Pt nanowire structures with larger aspect ratios have shown evidence of interesting coherence effects.  These include strong polarization of emitted light\cite{IngvarssonetAl07OE} and pronounced interference effect\cite{KleinetAl08APL}.  We see no evidence for such effects in our structures over the wavelength range examined, with no detectable net polarization of the emitted radiation.  This difference likely arises in part because the nanoconstrictions act more like point-like emitters than line-like emitters due to their highly nonuniform temperature profiles.

By standard optical measurement techniques we have used emitted blackbody radiation to characterize the temperature distribution in Pt nanowires before and during the electromigration process.  This approach requires no surface treatments or ultra high vacuum equipment, and with suitable infrared optics and detectors may be extended to lower temperatures.  We find that the maximum temperatures obtained in Pt nanowires can exceed the decomposition threshold of many organic compounds, reinforcing that care must be taken when using electromigration with molecules in place to fabricate structures for single-molecule electronic measurements.

The authors thank Peter Nordlander for useful discussions.  DN and DW acknowledge support from Lockheed Martin Corporation.  DN also acknowledges the David and Lucille Packard Foundation.  NH acknowledges support from Robert A. Welch Foundation grant C-1220.




\end{document}